# Composition dependence of electronic defects in $CuGaS_2$

Damilola Adeleye*, Mohit Sood, Michele Melchiorre, Alice Debot, Susanne Siebentritt

*Laboratory for Photovoltaics, Department of Physics and Materials Science, University of Luxembourg, Belvaux, L-4422, Luxembourg*

*Corresponding author: damilola.adeleye@uni.lu

## Abstract

$CuGaS_2$ films grown by physical vapour deposition have been studied by photoluminescence (PL) spectroscopy, using excitation intensity and temperature dependent analyses. We observe free and bound exciton recombinations, three donor-to-acceptor (DA) transitions, and deep-level transitions. The DA transitions at ~ 2.41 eV, 2.398 eV and ~ 2.29 eV are attributed to a common donor level ~ $38\pm5$ meV and three shallow acceptors at ~ 76 meV, ~ 90 meV and 210 meV above the valence band. This electronic structure is very similar to other chalcopyrite materials. The donor-acceptor transitions are accompanied by phonon replicas. Cu-rich and near-stoichiometric material is dominated by the transitions due to the acceptor at 210 meV. All films show deep-level transitions at ~ 2.15 eV and 1.85 eV due to broad deep defect bands. Slightly Cu-deficient films are dominated by intense transitions at ~ 2.45 eV, attributed to excitonic transitions and the broad defect transition at 2.15 eV.

# Introduction

$Cu(In,Ga)S_2$ (CIGS) is not only a promising material for a single junction solar cell, but also a strong candidate as a top cell in tandem applications to absorb high-energy photons of the solar spectrum [1-3]. An efficiency of 16% has been reported by Barreau et al. [4], considerably lower than the efficiencies of the selenide chalcopyrites $Cu(In,Ga)Se_2$ which have reached 23.6% (M. Edoff et al, in preparation) [5]. In particular, $Cu(In,Ga)S_2$ suffers from a high deficit in $V_{OC}$ [6-10]. This deficit is partly due to interface recombination, which can be mitigated by the correct choice of buffer layer, but also to a large part due to non-radiative recombination in the absorber bulk [10]. It is therefore essential to study the electronic defect structure of this semiconductor. The selenide chalcopyrite Cu(In,Ga)Se$_2$ together with the ternaries CuInSe$_2$ and CuGaSe$_2$ have been intensely studied and an understanding of the electronic structure and of the impact of composition has been established [11-18]. Accordingly, shallow donor and acceptor levels, as well as deep defects have been identified in both CuInSe$_2$ and CuGaSe$_2$. It is particularly interesting to compare the wide gap compound CuGaSe$_2$ with the low gap material CuInSe$_2$. For the wide bandgap CuGaSe$_2$, Spindler et al. have reported that defects levels shift mid-gap and defects that were shallow in CuInSe$_2$ become deeper in CuGaSe$_2$ [14]. Thus, as Ga is substituted for In, shallow defects become deeper and hence form deep levels which serve as channels for unwanted nonradiative recombination in Cu(In,Ga)Se$_2$ absorbers [11-14, 19, 20]. The electronic defects in Cu(In,Ga)S$_2$, unlike the selenide counterpart, are less studied in comparison to the selenide chalcopyrite [21-25].

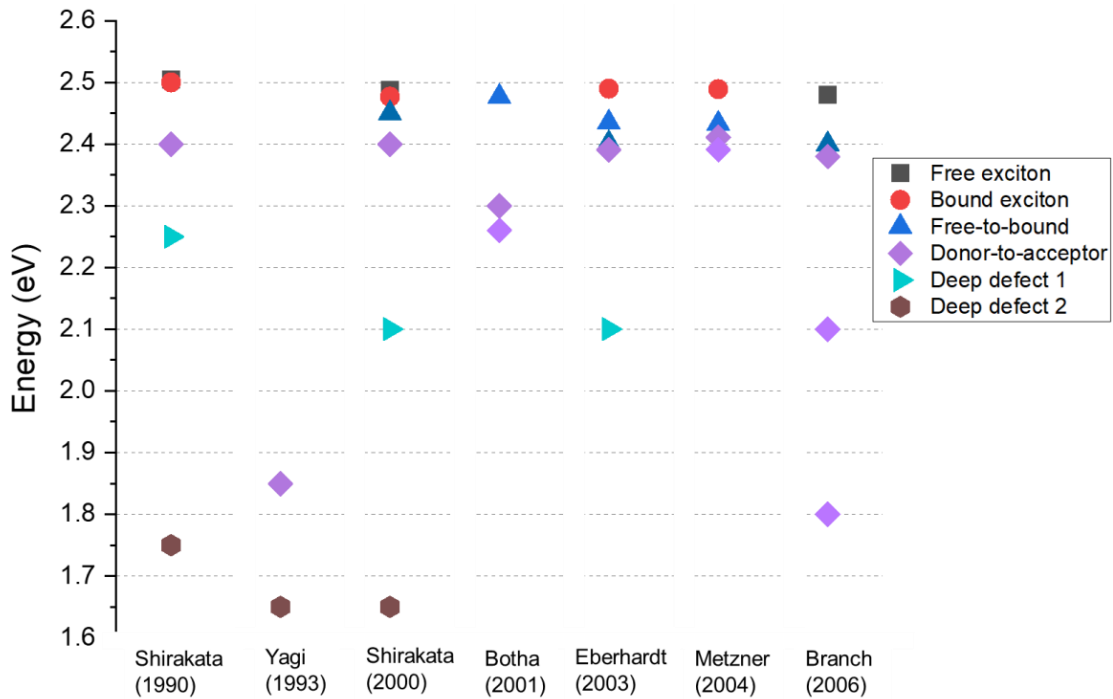

Figure 1: Overview of transition energies of CuGaS$_2$ identified from some preview studies.

However, it has been shown that the shallow defects in ternary $CuInS_2$ are similar to those in selenide chalcopyrites, with three shallow acceptors and one shallow donor, plus two deep broad defect bands close to 0.8 eV and 1.1 eV [9, 26]. For $CuGaS_2$, a comprehensive electronic defect structure is still incomplete [27-34]. For high efficiency sulfide chalcopyrite solar cells, it is necessary to include Ga [9], it is therefore essential to study the defect structure of $CuGaS_2$.

Previous reports on defects in $CuGaS_2$ have identified one or two donor-acceptor (DA) transitions around 2.39-2.41 eV, and these transitions were attributed to a common shallow donor around 20-25 meV [30, 35]. Earlier studies also reported a shallow donor energy level at 45-50 meV [27, 28]. In addition, deep level transitions are observed at 2.1 eV and around 1.7 eV to 1.8 eV [32, 36]. An additional deep transition around 2.3 eV has been identified as either a DA transition or due to a deep defect. The review of the photoluminescence transitions of $CuGaS_2$ from literature has been summarized in Fig. 1.

In this work, photoluminescence spectroscopy has been performed on $CuGaS_2$ films grown by physical vapour deposition to understand the electronic defect structure. This report will conclude by presenting a novel solar cell on Cu-rich $CuGaS_2$ absorber. A deeper understanding of the electronic defect structure in $CuGaS_2$, will also enhance the understanding of the role of Ga in $Cu(In,Ga)S_2$ films and solar cells.

## Deposition process for $CuGaS_2$ films

The polycrystalline $CuGaS_2$ films investigated in this work were deposited by one-stage co-evaporation of elemental copper and gallium with a source temperature of ~1250°C and ~1150°C, respectively, under a sulfur pressure between $5.9 \times 10^{-5}$ mbar to $8.5 \times 10^{-5}$ mbar. The various compositions of the $CuGaS_2$ films were obtained by changing the temperatures of the elemental sources and thus the flux of Ga and Cu. The deposition was on a molybdenum coated high temperature glass with better heat resistance than soda-lime glass [37] at actual substrate temperature of ~ 690°C. Such high substrate temperature is necessary to obtain high-quality Ga-containing films particularly for pure $CuGaS_2$ [38-40]. This is partly due to: (i) the slow elemental migration and reaction of Ga relative to In, and the relatively high melting point of Ga-based samples than In-based samples [38, 41, 42], as seen in the $Cu_2S$-$In_2S_3$ and $Cu_2S$-$Ga_2S_3$ phase diagrams [38, 42]. (ii) The higher melting point of sulfides compared to selenides due to the lower atomic weight of S compared to Se [43-45].

The characterization of the crystallinity, phases and vibrational properties of the films was by X-ray diffraction (XRD) using the CuKα radiation, and Raman spectroscopy with an excitation wavelength of 532 nm. The surface morphology and cross-section micrographs were obtained by a scanning electron microscope (SEM), and the chemical composition was determined by energy dispersive X-ray spectroscopy (EDX) with beam energy of 20 kV on as-grown films before etching. Therefore, the compositional ratio of Cu-rich films mentioned in this report is an integration of the ternary chalcopyrite phase and secondary copper sulfide ($Cu_xS$) phase. As such, "Cu-poor" refers to material with a ratio [Cu]/[Ga] ≤ 1, while "Cu-rich" refers to [Cu]/[Ga] ≥ 1. The Cu-excess phases were removed by etching in an aqueous solution of 10 % potassium cyanide (KCN) for 5 minutes [46], before the photoluminescence measurements.

Lower substrate temperatures during the deposition of the $CuGaS_2$ films resulted in poor quality films showing unidentifiable crystallographic phases among those close to $CuGa_3S_5$ and $CuGa_5S_8$ [47-49]. This shares a similarity to $CuGaSe_2$ when deposited at low temperatures [50]. Conversely, at these high deposition temperatures, group VI elements such as sulfur and selenium, have low sticking coefficient and are extremely volatile, which increases the possibility and the rate of sulfur loss and re-evaporation [34, 51-53], necessitating a high pressure of sulfur during growth. The growth parameters of films grown at various sulfur vapour pressures and substrate temperatures, labeled G1-G4, are presented in Table 1, the X-ray diffractogram on the films are also shown in Fig. 2.

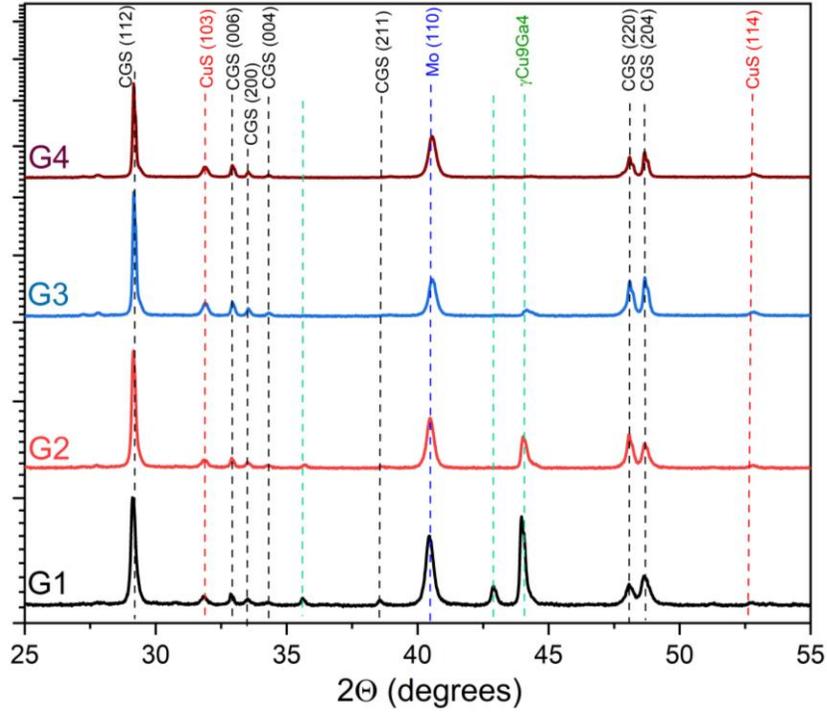

Figure 2: X-ray diffractogram of various films showing the effect of increasing growth temperatures and sulfur pressures from G1-G4. The ICDD PDF 00-025-0279 and 00-037- 1492 databases have been used to reference the peaks. The details are presented in Table 1.

Analysis of the chemical composition of film G1, deposited with actual substrate temperature of 600° C and chamber pressure of $5.9 \times 10^{-5}$ mbar, showed that the [S]/([Cu]+Ga) ratio was 0.69. From the X-ray diffractogram in Fig. 2, the deficiency of sulfur in G1 promoted the growth of a γCu9Ga4 phase. An increase in the growth temperature to 620 °C in G2 minutely increased the S content to 0.71 and the slight decrease of the γCu9Ga4 phase. Ultimately, by simultaneously increasing both the deposition temperature and chamber pressure to ~ 690°C and $5.9 \times 10^{-5}$ mbar respectively, the S content increased to 1.0 and the unwanted γCu9Ga4 phase was suppressed. Hence, the deposition of $CuGaS_2$ requires a larger S overpressure [27], than would be needed for pure $CuInS_2$ or $Cu(In,Ga)S_2$ [39, 54], in order to mitigate sulfur loss. Consequently, during the $CuGaS_2$ deposition process, the sulfur pressure in the chamber is maintained in the range of $5.9 \times 10^{-5}$ mbar to $8.5 \times 10^{-5}$ mbar. The deposition time of ~ 2 hours is used to achieve thicknesses of approximately 2μm.

Table 1: Influence of deposition properties on the sulfur-content in CuGaS$_2$ films

| Film label | Actual substrate temperature | Sulfur pressure | [S]/([Cu]+[Ga]) |
|---|---|---|---|
| **G1** | 600 °C | $3.7 \cdot 10^{-5}$ mbar | 0.69 |
| **G2** | 620 °C | $3.7 \cdot 10^{-5}$ mbar | 0.71 |
| **G3** | 640 °C | $4.7 \cdot 10^{-5}$ mbar | 0.81 |
| **G4** | 690° C | $5.9 \cdot 10^{-5}$ mbar | 1.0 |

## Effect of growth conditions on the structural properties of the films

Before presenting the results and discussing the optical characterization of different spectral regions on the films, it is imperative to ascertain the quality of the films under investigation. Hence, in the following section, the material characterization in terms of composition analyses, preferential chalcopyrite orientation, crystallinity and microstructural structural properties obtained from SEM-EDX, XRD and Raman analyses will be examined.

The chemical composition of the different films studied, as analyzed by EDX, is between 0.94 and 2.0 in [Cu]/[Ga] atomic ratio. Fig. 3 shows the SEM micrographs depicting the typical surface morphology and cross-sectional images of Cu-rich and Cu-poor films. The specific composition of the films shown are [Cu]/[Ga] = 1.3 for the Cu-rich film, and [Cu]/[Ga] = 0.94 for the Cu-poor film. The micrographs were obtained after the Cu$_x$S secondary phase was etched by 10 % KCN solution. The top-view, Fig. 3b, and the cross-sectional image, Fig. 3d of the Cu-poor film show rough granular surface with pyramidal grains which are compact and well-connected to the back on the Mo back-contact. On the other hand, both micrographs in Fig. 3a and Fig. 3c show that the Cu-rich films featured smoother surfaces with larger and denser grains. This is in accordance with other chalcopyrite compounds, where it is well established that copper-excess promotes the formation of large grain sizes and improves crystallinity [55-59]. Additionally, the high deposition temperature and pressure could have contributed to the quality of both Cu-rich and Cu-poor films, as these conditions foster effective nucleation and improve the quality of the grain growth [34, 37].

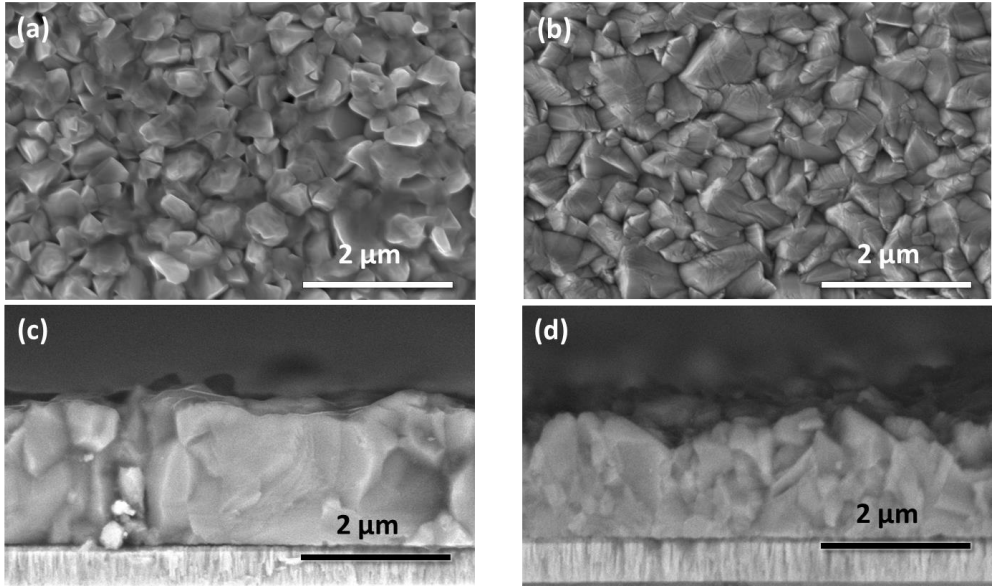

Figure 3: SEM micrograph showing the microstructure of CuGaS$_2$ films. Top-view and topographical view of typical (etched) Cu-rich film ((a) and (c)), and Cu-poor film ((b) and (d)).

The characteristic crystallographic orientation of the prevalent phase in the layers obtained from XRD is depicted in the diffractogram in Fig. 4a. The ICDD database - ICDD PDF 00-025-0279 is used as a reference to index the peaks. The most prominent peak is the (112) plane of CuGaS$_2$. The peak at 41° in Fig. 4a is due to the Mo back contact. A measure of the crystal quality is manifested in the split of the 220 and 204 peaks, resulting from the tetragonal distortion occurring in the chalcopyrite unit cell. The crystal quality of the films investigated is also corroborated by the absence of secondary phases in Fig. 4a, besides the CuS that is expected in a Cu-rich film. Fig. 4b shows the Raman spectrum of a Cu-rich CuGaS$_2$ film. The dominant line at 310 cm$^{-1}$ is the A1 mode, which corresponds to the vibration of the sulfur (or group VI) atom [60, 61]. This mode is also the dominant Raman mode in other chalcopyrite compounds such as CuInS$_2$, CuInSe$_2$, CuGaSe$_2$, etc. [62, 63].

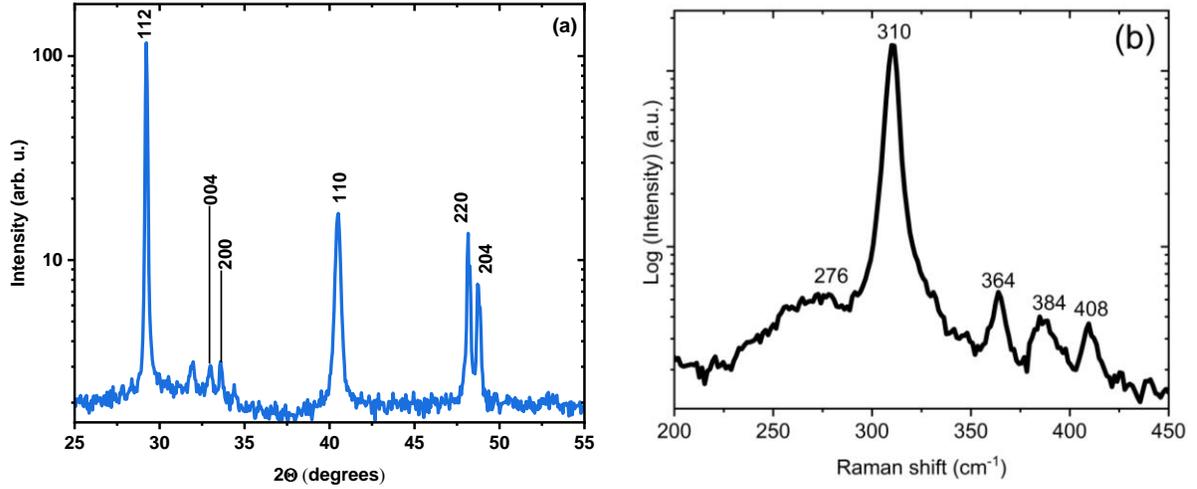

Figure 4: (a). X-ray diffractogram of the Cu-rich as-grown CuGaS$_2$ film in Fig. 3 showing the reflection planes. (b) Raman spectrum of a Cu-rich CuGaS$_2$ film with the Raman active modes.

The other less intense but notable Raman-active modes appearing at 276 cm$^{-1}$, 364 cm$^{-1}$ and 384cm$^{-1}$ correspond to the highest longitudinal optical phonon modes [60], while the peak at 408 cm$^{-1}$ has been attributed to MoS$_2$ [64]. Since Raman spectroscopy is surface-sensitive [65], the detection of the MoS2 peak could be due to holes in the film. The impact of the modes will be revisited in relation to the observed phonon replicas in the PL spectra discussed in the subsequent section. However, the nonappearance of any characteristic secondary phase, is an indication of the high quality of the CuGaS$_2$ film. To summarize, the results from the SEM-EDX, XRD and Raman analyses on the CuGaS$_2$ thin films investigated attest to the good quality of the films.

## Photoluminescence features of CuGaS$_2$ at low temperature

First, a summary of the PL spectra of CuGaS$_2$ with varying compositional ratios is presented together with the attributions for the different observed transitions. Afterwards, the methods used in analyzing and assigning the different peaks to a specific transition are discussed. Fig. 5 shows an overview of different CuGaS$_2$ PL spectra by composition at 10 K. The spectra feature (i) near band edge emissions with sharp intense excitonic (EXC) peaks around 2.48 eV, 2.49 eV and 2.502 eV; (ii) shallow defect-related emissions between 2.25 eV and 2.45 eV: several free-to-bound (FB) and donor-acceptor transitions (DA) with their phonon replicas, and (iii) a broad deep defect peak at ~ 2.15 eV. The influence of the [Cu]:[Ga] composition on some peaks can be clearly observed in the 2.3 eV transition (DA3), where the intensity of the peak increases with increasing Cu content, even dominating and screening other peaks in the spectrum for the film with [Cu]/[Ga] ratio = 2.

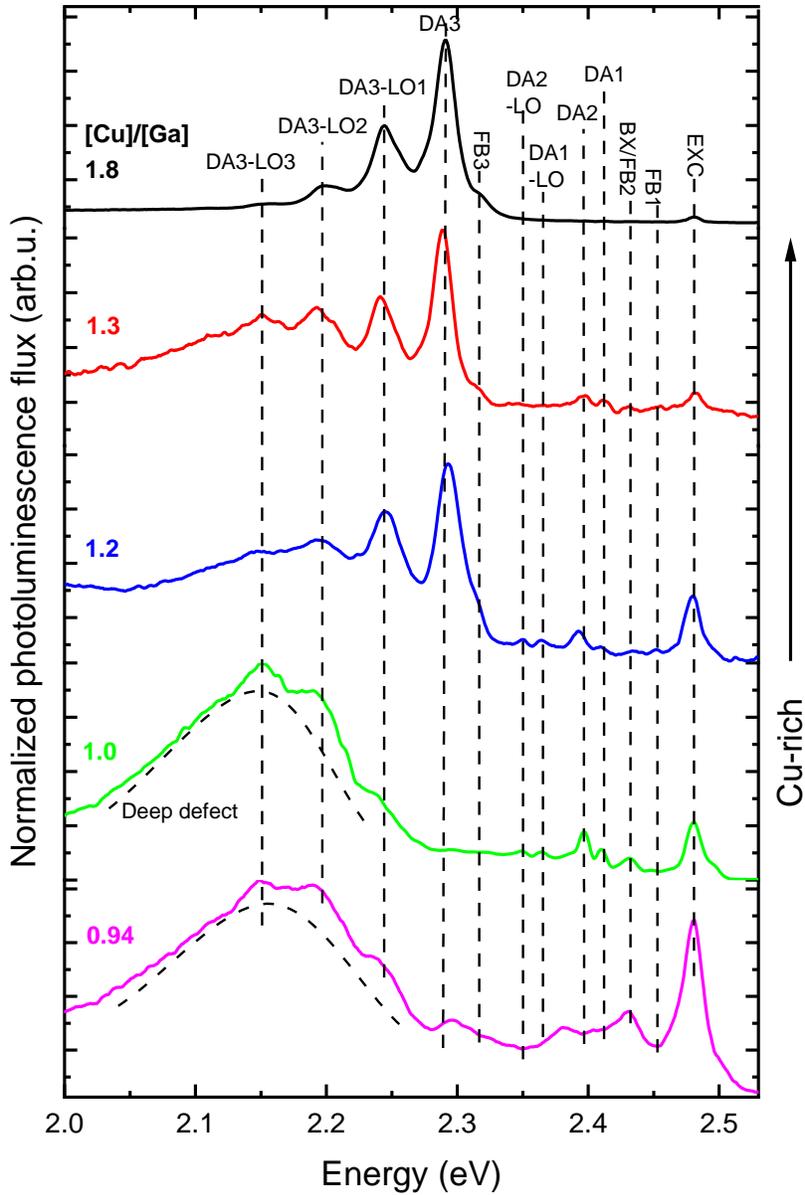

Figure 5: Photoluminescence spectra of $CuGaS_2$ with different chemical compositions at 10 K. The attribution of the peaks to the transitions in the figure will be derived in the following sections.

For slightly Cu-rich films with [Cu]/[Ga] ratio of 1.3 for example, the relative intensity of the 2.3 eV transition to the other peaks reduces, and it is noticeable that the 2.3 eV transition overlaps with the broad peak around 2.15 eV. In contrast, the intensity of the broad peak at ~ 2.15 eV and another one at 1.85 eV (see Fig. 17) increases with lower Cu-content and it dominates the Cu-poor material alongside the excitonic transition at 2.48 eV and transitions around 2.40 eV.

To investigate the different spectral regions, the relative intensity of the transitions described above is considered, as such, the near band-edge emissions and shallow defects are investigated using the near stoichiometric and Cu-rich films, while the deep defects are studied with Cu-poor and near stoichiometric films. The assignment of a peak to a specific transition follows the evaluation of PL flux and energy position in dependence of the excitation intensity in a double logarithmic scale and semi-logarithmic scale, respectively [66]. The high luminescence of some samples allowed for a wide range of excitation intensity over many orders of magnitude, and the double-logarithmic plot of the excitation-dependent integrated PL flux of such samples results in a curvature which cannot be described by a single power law.

The curvature is inherent and occurs when multiple defect levels participate in the recombination process [67]. Using rate equations and charge balance, exhaustive conditions beyond just the simple case, where a single power law can describe PL flux dependence on excitation intensity, has been reported in Ref [67-69]. A more comprehensive double-power law expression that better describes the curved shape

$$I_{PL} \propto \frac{\phi^{k_1+k_2}}{1+\left(\frac{\phi}{\phi_0}\right)^{k_1}} \tag{1}$$

where $k_i$, ($i$ = 1, 2) take on multiples of $\frac{1}{2}$ and $\phi_0$ is a turning point or crossover excitation at which a state interacting with the recombination process becomes saturated [67]. Essentially, for a curved double-log plot, the k-values for exciton-related transitions are between $\frac{2}{2} \leq k \leq \frac{4}{2}$, whereas for defect-related transitions $\frac{1}{2} \leq k \leq \frac{3}{2}$ [67]. A simple summary of *k*-value for different transitions investigated in this work is presented in Table 2, however, more complex cases can be found in Ref. [67, 68].

Table 2: Summary of the behaviour of the power law exponent ($k$) and $\beta$-values in dependence of excitation intensity. The values of $k$ take on multiples of $\frac{1}{2}$.

| Transition | Power law exponent ($k$) | | Change of energy position ($\beta$-value) |
|---|---|---|---|
| | Low ($\phi$) | High ($\phi$) | |
| Exciton | $\frac{4}{2}$ | $\frac{2}{2}$ | 0 |
| Donor-to-acceptor (DA) | $\frac{3}{2}$ | $\frac{1}{2}$ | 1-5 meV/decade |
| Free-to-bound (FB) | $\frac{3}{2}$ | $\frac{1}{2}$ | 0 |

To distinguish between DA and FB transitions we use the characteristic blue-shift of the emission energy of DA transition with increasing excitation intensity [66, 70]. This energy position is expressed by

$$E_{DA} = E_G - (E_D + E_A) + \frac{q^2}{4\pi\varepsilon_0\varepsilon_r R_{DA}} \tag{2}$$

where $E_{DA}$ is the DA peak energy position, $E_G$ is the bandgap, $E_D$ is the donor defect energy relative to the conduction band and $E_A$ is the acceptor defect energy relative to the valence band. The last term is the Coulomb energy, with $q$ being the elementary charge, $\varepsilon_0$ is vacuum permittivity, $\varepsilon_r$ is the relative permittivity and $R_{DA}$ is the spatial distance between the donor and acceptor [66, 71]. As the excitation intensity increases, the density of neutralized donors and acceptors increases, and the spatial distance $R_{DA}$ between the donor and acceptor atoms decreases, thereby increasing the influence of the Coulomb interaction. The relationship between the transition energy position in dependence of excitation is empirically described by

$$E_{DA}(\phi) \propto E_{DA}(\phi_0) + \beta \log\left(\frac{\phi}{\phi_0}\right) \tag{3}$$

where $\beta$ typically takes values between 1-5 meV per decade of excitation intensity [72].

## Near band-edge luminescence of CuGaS$_2$ (2.46-2.53 eV)

The band-edge emissions are assessed using the film with the highest Cu-content of [Cu]/[Ga] ratio = 2.0, due to its high luminescence flux and well-resolved peaks – although not obvious in Fig. 5 because of the high luminescence of the DA3 transition. The luminescence strength of this film also supports the enhanced crystallinity when the material is processed under high Cu excess. A plot of the PL spectra in the near band-edge region between 2.46 eV and 2.53 eV at different excitation intensities is illustrated in Fig. 6. We will argue in the following that the emission line B is the ground state of the free exciton, while A is the first excited state. The lines C-F will be identified as bound excitons.

Of the six peaks delineated, the most intense peak is at ~ 2.481 eV (D), with transitions at 2.488 eV (C) and 2.496 eV (B) at lower intensities, but visible in all spectra. On the high-energy end, the weak line at ~ 2.518 eV (A) is visible only at high excitation, and on the low-energy end, the intensity of a transition at ~ 2.474 eV (E) decreases while the 2.468 eV (F) peak is more resolvable at higher excitation intensities. In Fig. 6, the lines do not show a shift of energy position with increasing excitation intensity, which preliminarily leaves them as either excitonic or free-to-bound transitions. To discriminate between the two possibilities, the PL flux in dependence of the excitation intensity for the different peaks is evaluated in a double-log plot shown in Fig. 7a. The multiple-power law in equation (1) is rather used to fit the curves, and the fit of the emission line 2.496 eV (B) is presented in Fig. 7b as an example.

The *k*-exponent results in ~ $\frac{3}{2}$ at low excitation intensity, and in ~ $\frac{2}{2}$ at high excitation intensity. As mentioned in the introduction to this section, the exponents take on multiples of $\frac{1}{2}$, and the change of exponent occurs when competing transitions or a defect involved in the transition saturates. The line B transition at 2.496 eV is attributed to the free exciton transition, since it occupies the highest energy position (apart from line A which is only detected at higher excitation intensity and will be discussed later). Bound excitons appear at lower energies due to larger binding energies of the exciton to defects [28, 66, 73]. Attributing line B to the free exciton is further substantiated by its subsequent use in deducing the free exciton binding energy ($E_x$) from the first excited state, as shown next.

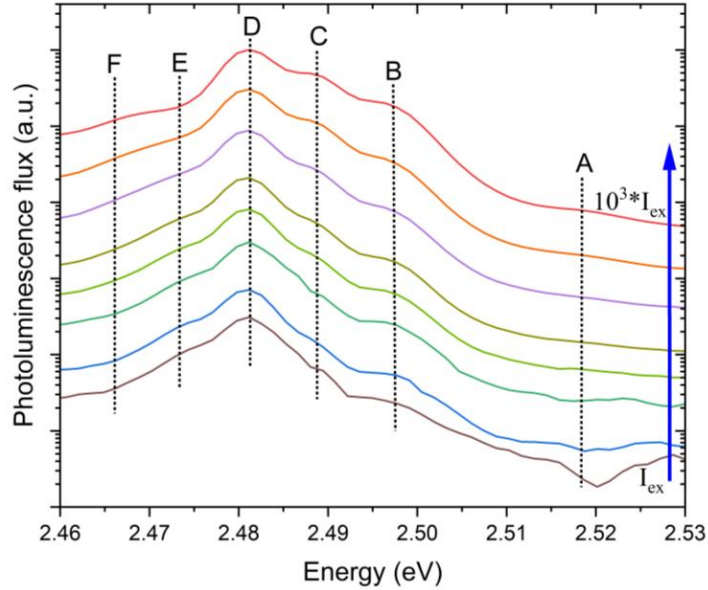

Figure 6: Near band-edge spectra of Cu-rich CuGaS$_2$ measured at 10 K taken at several excitation intensities over three orders of magnitude. The transition peaks are at 2.518 eV (A), 2.496 eV (B), 2.488 eV (C), 2.481 eV (D), ~2.474 eV (E) and 2.468 eV (F). The dashed lines highlight the constant energy positions with increasing excitation intensity.

In previous reports, and in agreement with this report, free exciton has been observed between 2.489 eV and 2.504 eV from photoreflectance spectroscopy and PL analyses [28, 31, 74]. In the different reports, the disparity of the free exciton energy positions were linked to lattice strain and the different analytical techniques [52]. The emission line A at 2.518 eV in Fig. 6 matches the first excited state ($n = 2$) of the free exciton, and the free exciton binding energy can be determined from the energy difference between the ground state $E_{FX}(n = 1)$ and the first excited state $E_{FX}(n = 2)$ using $E_x = \frac{4}{3}(E_{FX}(n = 2) - E_{FX}(n = 1))$. The determined free exciton binding energy of 29 meV is in the range of reported free exciton binding energy for CuGaS$_2$ between 28-32 meV [28, 36, 75], which further justifies the designation of line A as the first excited state of the free exciton. The knowledge of $E_X$ makes it possible to deduce the bandgap value at 10K, which will be important in the determination of the defect level energies. Therefore, in this study, we report the corresponding bandgap, ($E_g = E_{FX} + E_X$), for CuGaS$_2$ as 2.525 eV at 10 K.

For CuGaS$_2$, the hole effective mass ($m_h$) deduced from Hall-effect analysis and by calculation is $m_h = 0.69\, m_o$ [76, 77], where $m_o$ is the electron mass, and the dielectric constant obtained from optical-absorption analysis is $\varepsilon_o =$ 8.5 [78]. Different values between 0.12-0.19$m_o$ have been reported for the reduced mass of CuGaS$_2$ by different groups, consequently, the electron effective mass ($m_e$) deduced from the reduced mass is between 0.13-0.26 $m_o$ [28, 45, 76, 77, 79]. Therefore, the mass ratio ($m_e/m_h$) for CuGaS$_2$ is between 0.19-0.38. Sharma et al. found that the limit of mass ratio for a stable exciton, bound to a charged donor and a charged acceptor is 0.20 and 0.29 respectively [80], as such, the mass ratio for CuGaS$_2$ suggests that the binding of excitons to both ionized donors and acceptors in CuGaS$_2$ would result in unstable ionized complexes [80].

However, binding energy for the neutral complex of both the donor ($D^0, X$) and the acceptor ($A^0, X$) can be found from the expressions

$$E_{(D^0,X)} = 0.12 E_D + E_X \tag{4}$$

$$E_{(A^0,X)} = 0.07 E_A + E_X \tag{5}$$

where, $E_D$ and $E_A$ are donor and acceptor energies respectively [28, 80, 81]. Similar to the deduction of the binding energy for free exciton, the difference between a bound exciton and the bandgap corresponds to the binding energy of the bound exciton [82, 83].

From the knowledge of the bandgap and exciton binding energy, the probable ionization energies of the donors or acceptor corresponding to an emission line can be calculated from equation (4) and equation (5). The values are summarized in Table 3 for emission lines C to F. Previous reports have associated similar transition to the line C at 2.488 eV to a bound exciton recombination [30, 56], while some other reports have attributed a comparable emission to the 2.481 eV line (D) as a FB recombination involving a transition between a neutral donor and the valence band-edge [56, 84]. According to the estimation presented in Table 3, it seems that the 2.488 eV exciton (C) is bound to neutral acceptor at 67 meV or neutral donor at 114 meV, while the 2.481 eV emission (D), is bound exciton to a neutral donor at 125 meV or a neutral acceptor 214 meV.

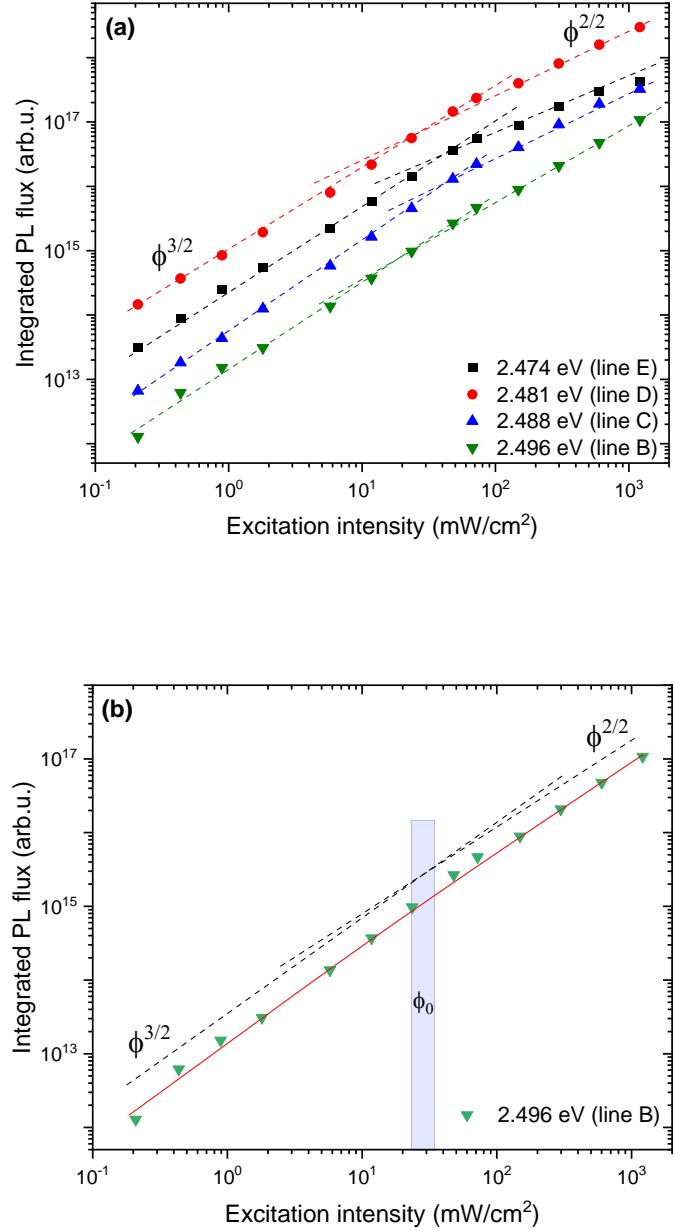

Figure 7: (a) Excitation intensity dependence of integrated PL flux for transitions lines 2.474 eV (E), 2.481 eV (D), 2.488eV (C) and 2.496 eV (B) fitted with the double power law. (b) A fit of the emission line 2.496 eV (B) with two power laws exponents $k = \frac{3}{2}$ and $k = \frac{3}{2}$ at high and low excitation, respectively. $\phi_0$ denotes the turning point between the two excitation regimes, i.e., the flux where one of the defects is saturated.

The existence of either of these levels and applicable level is presented in the succeeding sections. As we will show in the following, the only shallow donor we find has a binding energy of 35 meV, which makes it unlikely that any of those excitons are bound to a donor. On the other hand, we find shallow acceptor states at energies near 100 meV and 200 meV, to which the bound excitons C and D would correspond. Additionally, we find several deep defects, for which we can only speculate at the moment, they might be the defects to which the excitonic lines E and F are bound.

Table 3: Estimated values of exciton binding energies, neutral donor and acceptor energy levels calculated using equation (4) and equation (5) for the emission lines C-F.

| Line | Emission (eV) | Binding energy (meV) | Neutral donor $E_D$ (meV) | Neutral Acceptor $E_A$ (meV) |
|---|---|---|---|---|
| C | 2.488 | 37 | 67 | 114 |
| D | 2.481 | 44 | 125 | 214 |
| E | 2.474 | 51 | 183 | 314 |
| F | 2.468 | 57 | 233 | 400 |

Lastly, in previous reports, transitions identical to the line E have been assigned to a FB transition involving a shallow level [27, 28]. However, following the excitation-dependent analyses of the line E showing exciton-related behavior. The consideration of the transitions at 2.474 eV (line E) and 2.468 eV (line F) as exciton-related transitions would require that the exciton be bound to a deep defect level, as inferred from Table 3.

# Shallow defects, donor-to-acceptor pair transitions and phonon coupling

Several sharp peaks dominate the typical PL spectrum of Cu-rich $CuGaS_2$ at 10 K, between the range of 2.45 eV and 2.10 eV, as seen in Fig. 5 and Fig. 8. Some of the peaks appear in groups at regular energy intervals, and as it will be shown in the following: these are phonon replicas associated with shallow donor-to-acceptor (DA) transitions. The series of sharp peak follows an intense line, known as the zero-phonon line (ZPL), which is followed on its low energy end by several successive peaks of weakening intensities. These peaks are separated by the energy of the coupling LO-phonon. The excitation- and temperature-dependence behavior of the phonon-replicas is identical to the emission at the ZPL. As we show below, the spectral intensity dependence of such phonon-assisted transitions is well described by the Poisson distribution expressed by,

$$I_n \propto \exp(-S)\frac{S^n}{n!}, \qquad (6)$$

where $n$ is the number of phonons involved in the interaction, $I_n$ is the intensity of the $n$th phonon replica and $S$, known as the Huang-Rhys factor, is the coupling strength of the electron-phonon interaction of the corresponding defect [85]. For shallow (weakly localized) defects, the electron-phonon coupling is weak and $S < 1$, thus, the ZPL is the most intense peak and does not shift in peak energy. On the other hand, if $S = 1$, there is a change in the maximum intensity, as the first phonon replica becomes of the same intensity as the ZPL. Lastly, when $S > 1$ there is a strong electron-phonon coupling of localized defects, leading to a shift in the maximum intensity away from the ZPL to a lower energy, since the phonon replicas have higher intensities than the ZPL. It is worth mentioning that for broadened emission bands, phonon replicas do not manifest by the sharp peaks, rather by a broad asymmetric distribution [86-88].

In the next subsections, each of the donor-to-acceptor pair transitions (DA) peaks as shown in Fig. 8, that is, DA1, DA2 and DA3, along with their accompanying phonon replicas, will be discussed. Interestingly, the corresponding free-to-bound transitions between the conduction band and the acceptor level are already observed at low temperature.

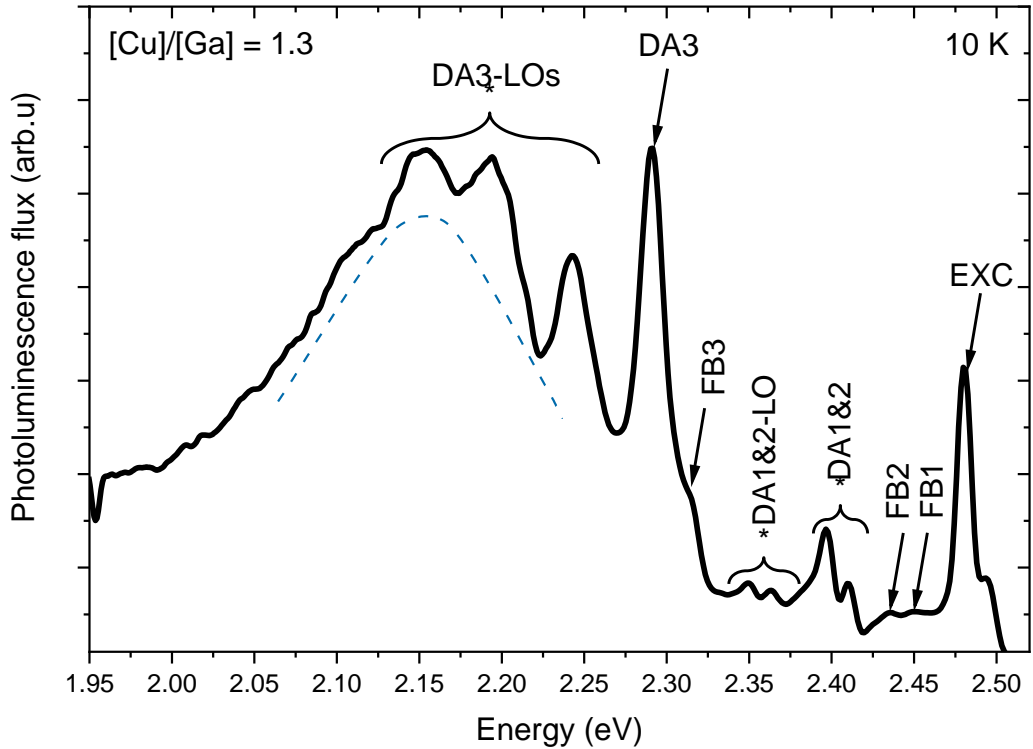

Figure 8: Low temperature (10 K) spectrum of Cu-rich CuGaS$_2$ film of [Cu]/[Ga] = 1.3 measured at 10 µW. Phonon replicas follow three DA transitions between 2.42 eV and 2.10 eV. This is the same film in Fig. 5 at a higher excitation intensity. The sample is chosen at this excitation power to highlight the appearance of several DA transitions and their replicas all together. The dashed line centered at approximately 2.15 eV delineates a broad transition around the said energy.

## DA3 transition at ~ 2.29 eV

The low-temperature (10 K) PL spectrum showing the transition related to 2.29 eV, measured at a low excitation intensity where the peaks are well resolved and without the strong influence of other defect peaks, is presented in Fig. 9. It is worth noting that the sample in Fig. 9 is the same as the sample in Fig. 5 with [Cu]/[Ga] = 1.8. However, while the spectrum presented in Fig. 9 is measured at 10 µW, the spectrum presented in Fig. 5 is at 100 µW. The spectrum (Fig. 9) features a series of peaks with the most intense line at ~ 2.29 eV followed by several successive lines of weakening intensities on the lower energy end. These weakening lines are energy-spaced by ~ 45±1 meV, corresponding to the lowest of the three highest energy optical phonon modes of 45.2 meV, 47.6 meV and 49 meV [60, 89], which are equivalent to the Raman modes observed at the frequencies of 364 cm$^{-1}$, 384 cm$^{-1}$ and 408 cm$^{-1}$ as seen in the Raman spectrum of CuGaS$_2$ presented in Fig. 4b.

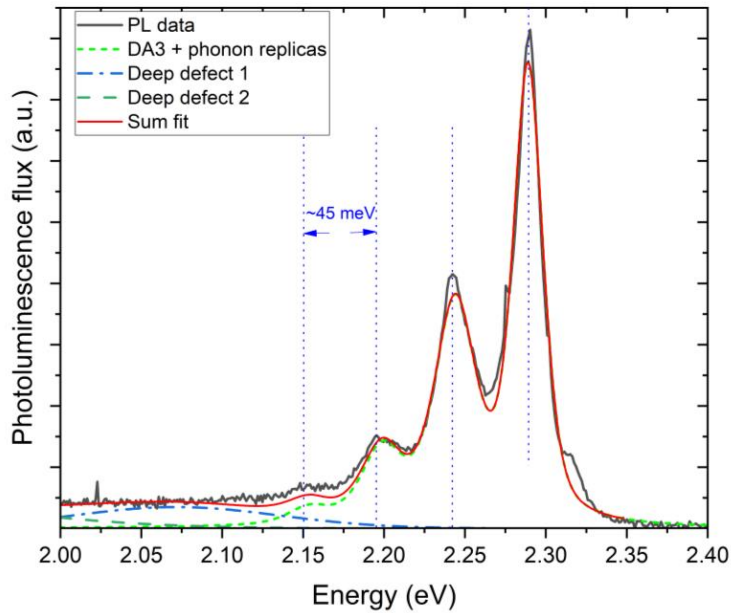

Figure 9: Low-temperature (10 K) PL spectrum of a CuGaS$_2$ film of [Cu]/[Ga] = 1.8 between 2.0-2.4 eV. The figure shows a fit of the phonon-assisted transition at ~ 2.29 eV (DA3) by Poisson function with consideration for deep defects. The low intense peak at ~ 2.32 eV is an associated FB transition related to DA3 which will be discussed afterwards.

A fit of the spectral and intensity pattern by the Poisson distribution in equation (6), while also considering a background of emissions from deep defects, yielded a Huang-Rhys factor $S \approx 0.80 \pm 0.05$ and a ZPL at ~ 2.285 eV. This value of S and the energetic distance between the ZPL and the band gap are in agreement: for defect transition more than 200meV away from the bandgap a rather high Huang-Rhys factor is expected [66, 87].

To identify the exact nature of the transition, the PL spectra acquired at different excitation intensities for the energy between 2.10-2.35 eV are presented in Fig. 10a. It is visible that there is a blue-shift of peak positions for the ZPL and the phonon replicas in parallel as the excitation intensity is increased. Such a shift in energy position is due to the influence of Coulomb interaction and indicative of a DA transition and it is expressed by the equation (3).

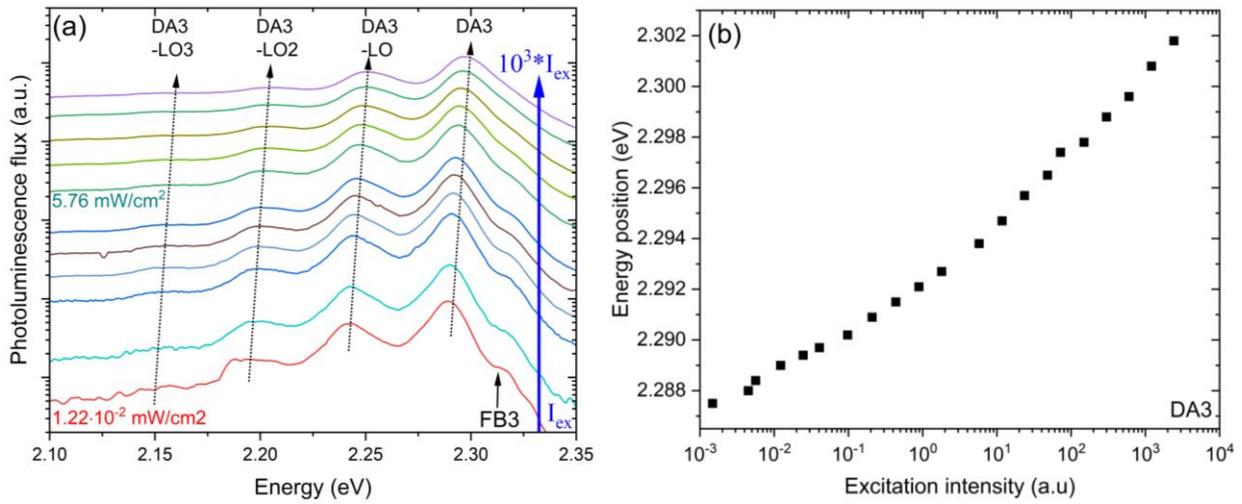

Figure 10: (a) The low-temperature (10 K) PL spectra of Cu-rich $CuGaS_2$ at different excitation intensity, demonstrating the shift of energy position of DA3 and its phonon replicas with the increase in excitation intensity. The dotted arrows are used to guide the eye for the shift in energy position. (b) Excitation intensity dependence of the energy position of DA3 transition in a semi-logarithm plot at 10 K.

The actual shift of energy position can be extracted from a plot of the energy positions against the excitation intensity. As shown in Fig. 10b, for the transition at ~ 2.29 eV, the plot of energy position against excitation intensity shows a curvature. This is due to the fact that, for a sufficiently wide range of excitation intensity, the energy positions of DA transitions assume an S-shape behaviour [67, 90]. The peak position approaches the energy position for infinite donor-acceptor pair separation at the lowest excitation, while at the highest excitation; the peak position approaches the summation of infinite donor-acceptor pair separation and the Coulomb energy for minimum donor-acceptor pair separation [90]. Excitation dependence of the integrated PL flux for the DA3 transition is reported in a double-logarithm plot shown in Fig. 11. It can be seen that the plot in Fig. 11 is a curvature which is adequately evaluated by equation (1). The fitting by two power law exponents results in $k = \frac{2}{2}$ at low excitation intensity and $k = \frac{1}{2}$ at higher excitation intensity. The change of exponents, referred to as crossover, occurs at ~ 3-6 mW/cm² of excitation intensity. This crossover indicates that a defect level or a deeper mid-gap level interacting with the recombination process of DA3 transition saturates at this intensity [67].

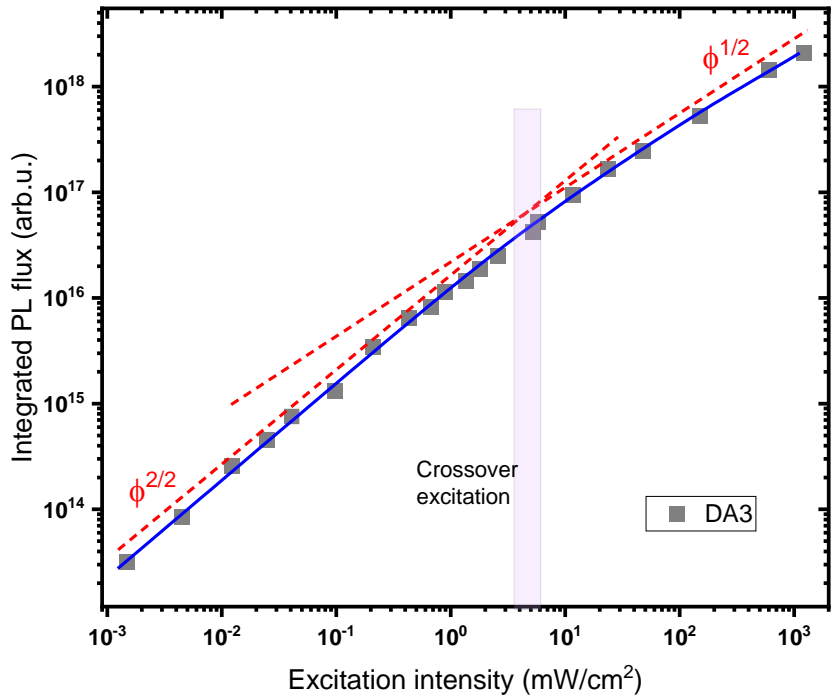

Figure 11: Double logarithmic plot of the DA3 transition with the integrated PL flux as a function of excitation intensity. The values are extracted from the integrated PL flux of the Cu-rich CuGaS$_2$ spectra in Fig. 10a.

On the high-energy end of DA3, is a low-intensity peak at ~ 2.32 eV as seen in Fig. 9 and Fig. 10a. The peak becomes more intense with increasing excitation intensity, as shown in Fig. 10a, until it is eventually obscured by the broadening DA3 transition. Nevertheless, it is still noticeable in Fig. 10a that the energy position barely changes with increasing excitation intensity. Given that the energy position of FB transition does not shift with energy position, and owing to its proximity to the DA3 transition, the weak peak at ~ 2.32 eV is assigned FB3. It is noteworthy that the FB3 transition might account for the curvature of the excitation dependence of PL flux for DA3 as illustrated in Fig. 11, since a shallow defect participating in the DA3 transition could saturate [67]. This is established by the value of the crossover excitation at ~ 3-6 mW/cm2 in Fig. 11 being close to the screening of FB3 in Fig. 10a, as seen above in the PL spectrum at 5.76 mW/cm$^2$ in Fig. 10a. Temperature-dependent analyses of the PL spectra shown in Fig. S2 of the Supplementary information, give further support to this attribution of DA3 and FB3. As the temperature increases, the intensity of DA3 decreases whereas the relative intensity of FB3 increases before the thermal quenching of the transition.

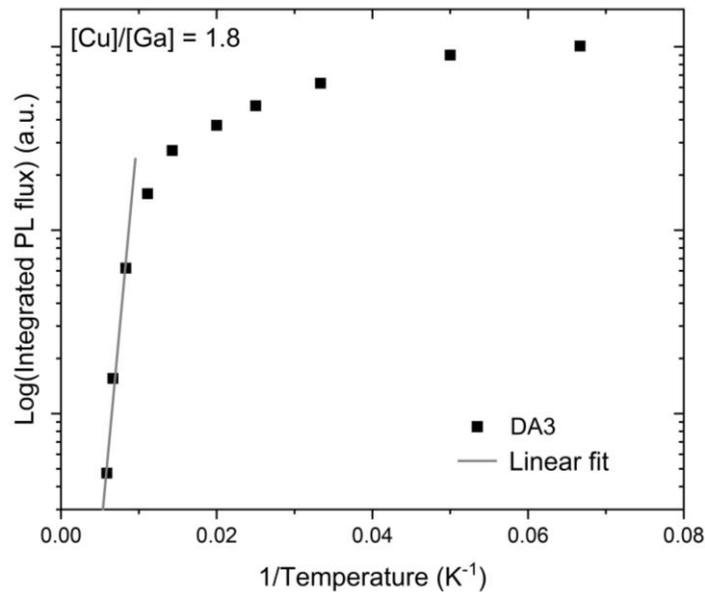

Figure 12: Arrhenius plot of the integrated PL flux with respect to temperature for the thermal quenching of the DA3 transition in a Cu-rich CuGaS$_2$ film.

This is another reason to associate this transition with FB3 because, ideally, shallow defect involved in a DA transition is thermally emptied with increasing temperature, hence leaving the FB transition. The Arrhenius plot of the integrated PL flux against the inverse temperature for the thermal quenching of the DA3 transition is shown in Fig. 12. The thermal activation energy was obtained as ~ 35 meV. This energy can be associated with the shallower defect involved in the DA transition, which is the donor, based on the lower effective mass of electrons. The energetic difference between FB3 and DA3, is compatible with the emptying for donor level at 35meV. With the knowledge the bandgap of 2.525 eV calculated for $CuGaS_2$ at 10 K, we derive the acceptor level of DA3 from the energy difference between FB3 and the bandgap as ~ 210 meV. As such, we conclude that DA3 is a transition between the donor level at 35 meV below the conduction band and an acceptor level at 210 meV above the valence.

## DA1 and DA2 transitions around 2.40 eV

As it will be shown in detail in the following, below the band-edge around 2.32-2.46 eV, two donor-to-acceptor transitions are identified at 2.410 eV and 2.398 eV, followed at an energetic distance of ~ 46 meV by peaks at 2.35 eV and 2.363 eV respectively, as shown in Fig. 13a. Additionally, on the high energy wing, two free-to-bound transitions were detected at ~ 2.43 eV and 2.45 eV. Fitting the donor-to-acceptor lines with the Poisson distribution determined the zero phonon lines to be at 2.410 eV (DA1) and 2.398 eV (DA2). The Huang-Ryhs factors were $S_{DA2} = 0.50 \pm 0.10$ and $S_{DA1} = 0.55 \pm 0.10$ for DA1 and DA2, respectively. These values are smaller in comparison to the $S$ factor for the DA3 as expected, since the deeper the defect level, the more localized and tightly bound the carriers are to the defects, hence an even stronger electron-phonon coupling [87].

The integrated PL flux in dependence of excitation intensity presented in Fig. 13b shows that both DA1 and DA2 can be fitted by a single power law yielding an exponent $k \approx \frac{2}{2}$. As mentioned in the preceding Section, $k$ takes on multiples of $\frac{1}{2}$, and for DA transitions in particular, $k$ approaches $\frac{2}{2}$ at low excitation intensities [67]. Additionally, the shift of energy position for both DA1 and DA2 is ~ 2.4 meV/decade as shown in Fig. 13c. Hence, it can be concluded that the study of DA1 and DA2 transition in this study is within the limiting region of low excitation.

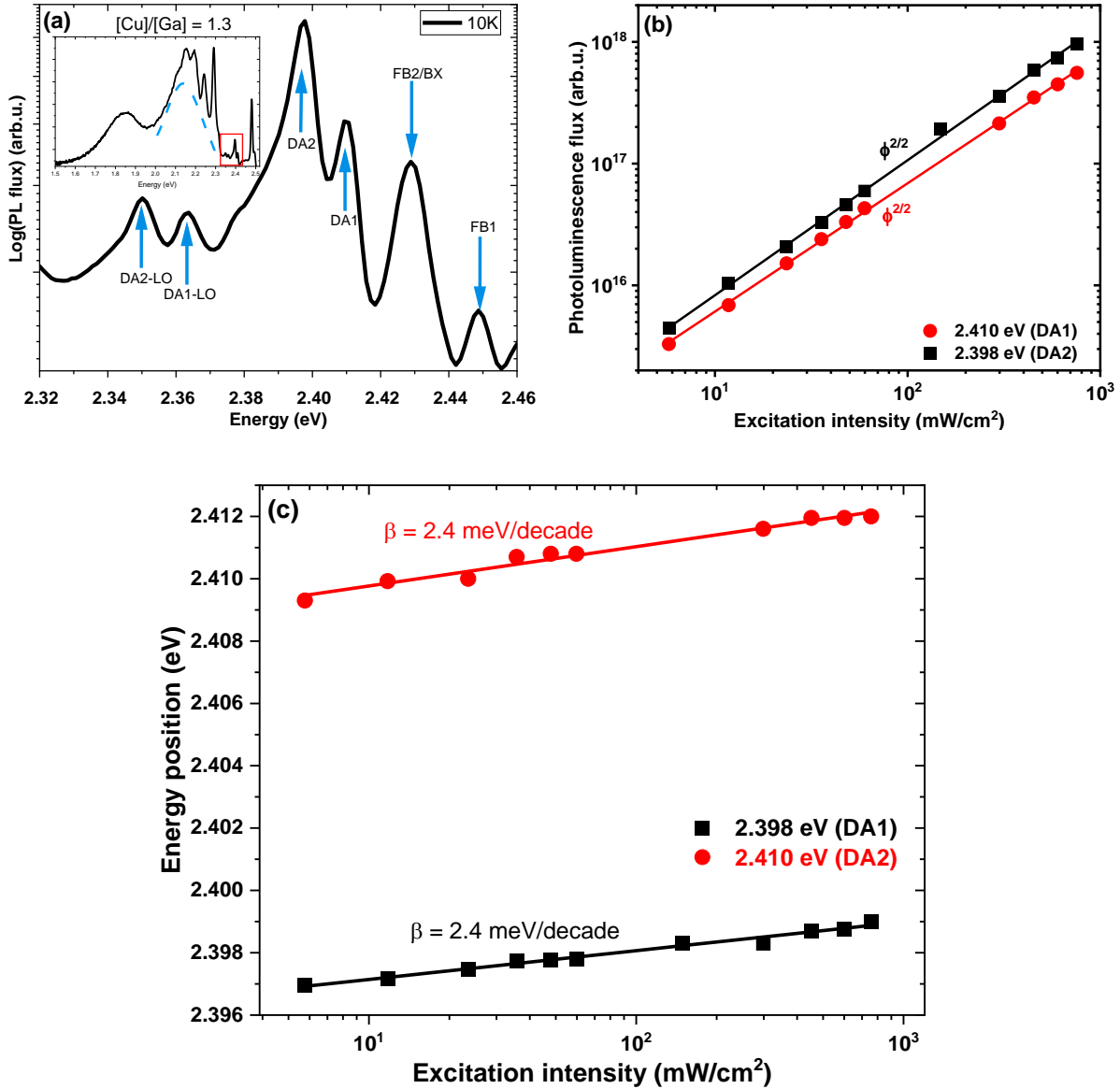

Figure 13: (a) Low temperature (10 K) PL spectrum of Cu-rich CuGaS$_2$ measured at 0.9 mW/cm$^2$, showing phonon replicas accompanying the DA transitions at 2.410 eV (DA1) and 2.398 eV (DA2). The sample used to analyze the transitions is the CuGaS$_2$ film with [Cu]/[Ga] = 1.3 in Fig. 5 and Fig. 8. In Inset is the full spectrum of the PL spectrum of the film from Fig. 8. The dashed line centered at approximately 2.15 eV describes a broad transition related to a deep defect. The region in focus is indicated by the red box in the inset. (b) The Integrated PL flux as a function excitation intensity of the 2.398 eV (DA2) and 2.410 eV (DA1) transitions at 300 K. (c) Excitation intensity dependence of the energy position of DA1 and DA2 transition in a semi-logarithm plot at 10 K.

For the transitions at ~ 2.43 eV and 2.45 eV indicated as FB2/BX and FB1 respectively in Fig. 13a, the integrated PL flux for both peaks with respect to excitation intensity in a double-log scale is shown in Fig. 14a. The single power law fit of both transitions also gives a power law exponent $k = \frac{2}{2}$ for both transitions. This linear dependence of the PL flux on excitation intensity can be interpreted as transitions originating from DA at low excitation, FB transitions or BX transition [66, 67], although both transitions at ~ 2.43 eV and 2.45 eV have been tentatively reported as FB transitions [30, 31]. The energy positions in dependence of excitation intensity which is presented in Fig. 14b show no significant shift of energy position with increasing excitation intensity over three orders of magnitude for the 2.45 eV peak, making its consideration as an FB transition compelling.

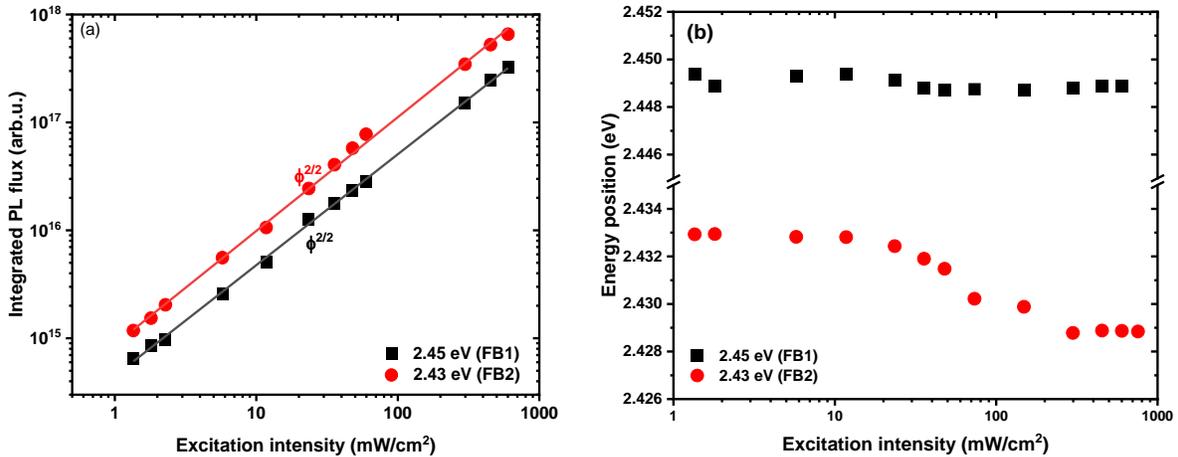

Figure 14: (a) PL flux in dependence of the excitation intensity for the transition peaks at 2.43 eV (FB2) and 2.45 eV (FB1). (b) Energy position in dependence of peak energy position for transitions at 2.43 eV and 2.45 eV.

The behaviour of the ~ 2.43 eV transition in Fig. 14b shows a shift of energy position, where the energy position is initially at ~ 2.433 eV later shifting to ~ 2.428 eV. The explanation for the energy shift of 2.43 eV peak is by the existence of two transitions occurring around 2.35 eV, i.e., ~ 2.428 eV and 2.436 eV, which are better resolved by the temperature-dependent analysis. Therefore, we conclude that below the excitation intensity of 10 mW/cm$^2$ in Fig. 14b, the 2.428 eV transition dominates, however, as the excitation intensity increases beyond 10 mW/cm$^2$, the intensity of the 2.436 eV transition increases and dominates, hence the shift of energy position observed in Fig. 14b.

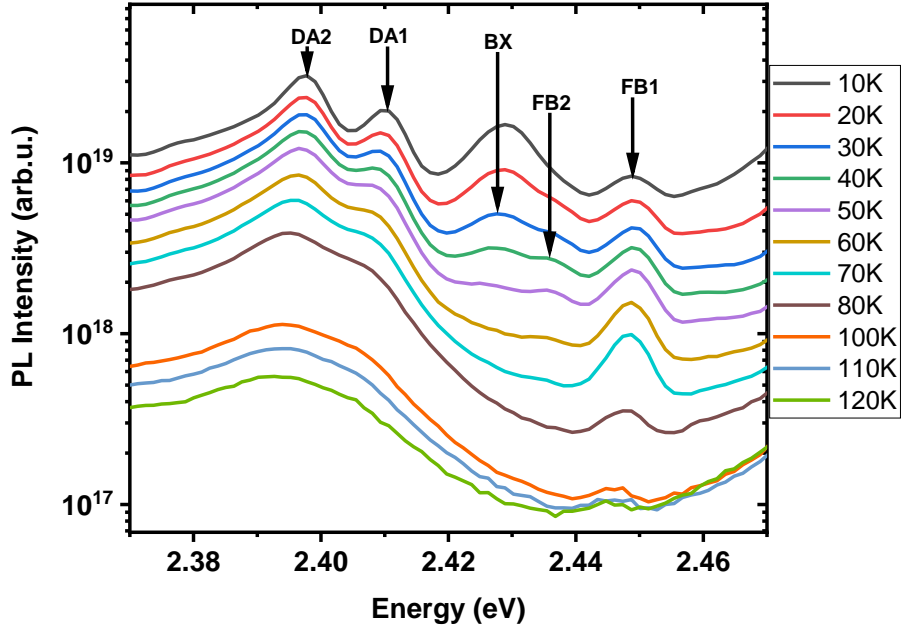

Figure 15: Temperature dependent PL spectra on Cu-rich film with [Cu]/[Ga] ratio of 1.3. The temperature-dependent measurement leads to the resolution of bound exciton transition at 2.428 eV and free-to-bound peaks at 2.436 eV and 2.449 eV.

Temperature-dependent measurements were performed to understand the behaviour of the DA1 and DA2 transitions, and to know the influence of temperature on the associated FB or excitonic transitions associated with DA1 and DA2. This is because shallow defects will be thermally emptied with increasing temperatures, and contribute to a FB transitions [66]. The temperature-dependent spectra presented in Fig. 15, show that as the temperature increases, the intensity of the DA1 and DA2 peaks decrease since a shallow defect level involved in the transitions is thermally emptied. It becomes obvious that the 2.43 eV line (labelled FB2/BX in Fig. 13a) consists of two peaks, one at 2.428 eV and another at 2.436 eV. It is observed that the relative intensity of the 2.428 eV line (BX) rapidly decreases and is quenched at approximately 50 K, as is typical for a bound exciton. Its energy position suggests that, it is bound to a much deeper defect than the excitons discussed above in section 4.

The relative intensity (compared to the DA transitions) of the 2.436 eV (FB2) and 2.448 eV (FB1) lines increases as temperature increases up to 70 K before decreasing and quenching at 120 K, supporting their attribution as FB transitions. Given the proximity to the energy positions of DA1 (2.410 eV) and DA2 (2.398 eV), the transitions at 2.449 eV and 2.436 eV can be sufficiently associated with the DA1 and DA2 transitions as the related FB transitions at, FB1 (2.449 eV) and FB2 (2.436 eV) respectively. By the energy difference between the DAs and FBs, FB1 and FB2 appear to involve a common shallow donor at 35 meV. In accordance with the attribution of FB1 and FB2, and the estimated $CuGaS_2$ bandgap of 2.525 eV at 10K, the 2.449 eV (FB1) and 2.436 eV (FB2) transitions are estimated to involve acceptor levels at ~ 75 meV and ~ 90 meV, respectively.

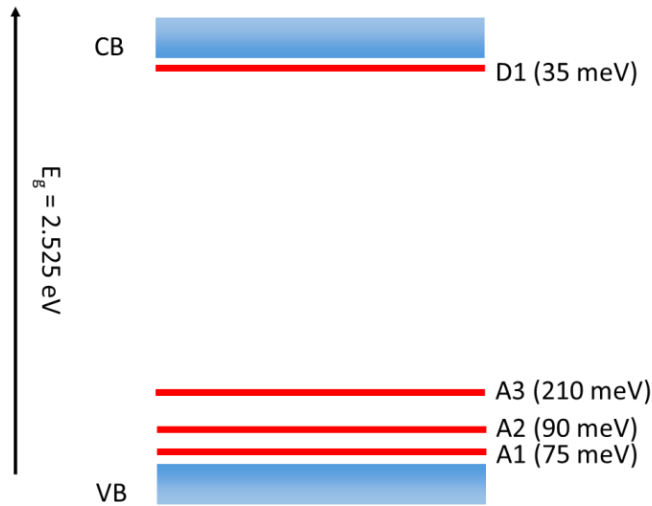

Figure 16: Tentative shallow defect levels in $CuGaS_2$ showing a donor level (D1) and three shallow acceptor levels (A1, A2 and A3).

At this point, in the analyses of the different transition peaks of $CuGaS_2$ studied in this report, a summary of the shallow defects in $CuGaS_2$ can already be drawn. The transitions DA1, DA2 and DA3 are assigned donor-to-acceptor transitions due to the blueshift of their energy positions with excitation ($\beta$-value) in dependence to excitation intensity, and by the nature of their power law exponents ($k$). For DA3, the power law exponent changed from $k = \frac{2}{2}$ to $k = \frac{1}{2}$, while $k$ is 1 for both DA1 and DA2. Summarily, there exist a common shallow donor (D1) level at ~ 35 meV and three shallow donor levels at 75 meV (A1), 90 meV (A2) and 210 meV (A3). The tentative shallow defect levels involved in DA1, DA2 and DA3 transitions are illustrated in Fig. 16.

## Deep defects at 2.15 eV and 1.85 eV

The PL spectra of all the films (in Fig. 5) investigated show that, as the compositional ratio of [Cu]/[Ga] decreases, the spectrum is dominated by two broad transitions in the range between 1.6 eV and 2.3 eV. The two transitions centered at 2.15 eV and 1.85 eV are as shown in Fig. 17a. The occurrence of the transition at 2.15 eV is strongly composition dependent, as shown by the reports of Botha et al. and Metzner et al. [33, 36]. Both groups report that in Ga-rich samples the transition shifts to ~ 2.0 eV, while in Cu-rich samples the transition shifts to higher energies around 2.12-2.18 eV. Hence it is possible that one or more defects are involved in the 2.15 eV transition, and this may account for the broadness of the peak. It is worth mentioning that in a Cu-poor film which is not presented in this report, it is possible to fit a peak at ~ 2.0 eV, hence it is possible that there is an additional transition at ~ 2.0 eV. However, given that a Cu-rich film is used to investigate this peak in this report, the transition centered at 2.15 eV is in agreement with previous reports.

Additionally, the phonon replicas on DA3 superpose on the 2.15 eV peak, thereby imposing constraints when fitting with an assumed Gaussian shape.

Analyses of the power law dependence for both deep transitions yield exponents $k \sim 1$. The energy position with respect to excitation intensity over four orders of magnitude show a blue-shift of ~ 5 meV per decade for the 2.15 eV peak, while the 1.85 eV peak shows a larger blue-shift of ~ 25 meV per decade with respect to excitation intensity as presented in Fig. 17b. While the 2.15 eV transition show features of a typical donor-acceptor transition, the blue-shift with 1.85 eV is much larger than expected for a shallow DA transition. Therefore, in the past, it has been attributed to potential fluctuations of the band-edges [36]. However, if the transition is broadened due to potential fluctuations, all transitions would be broadened in the same way. Therefore, potential fluctuations can be excluded as the source of the strong blue shift. Although it is apparent that both transitions likely involve broad density of states, the large blue-shift of the 1.85 eV transition might also be due to stronger phonon coupling manifested with deep defects.

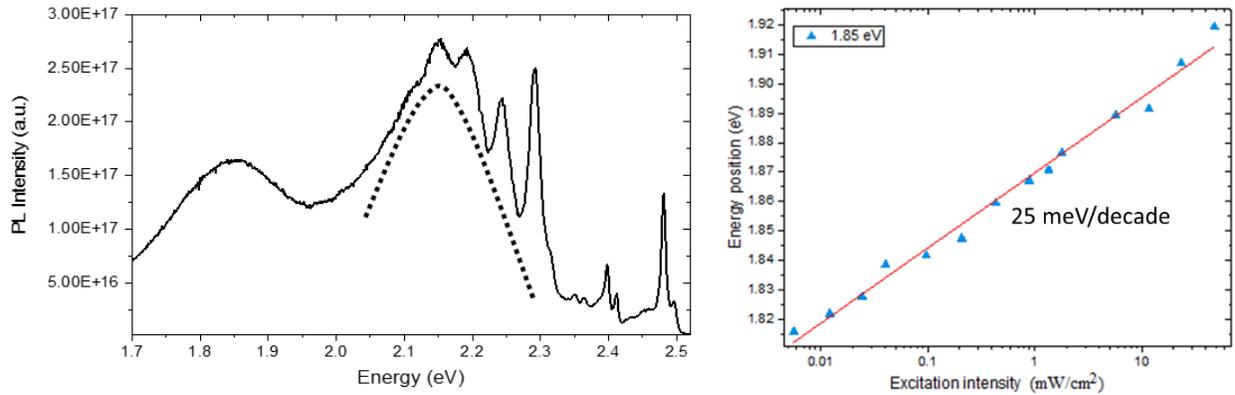

Figure 17: (a) Region of broadband deep defects featuring transitions centered at approximately 1.85 eV and 2.15 eV. (b) Energy position as a function of excitation intensity for the deep defect at ~ 1.85 eV.

## Summary and tentative shallow defect levels in CuGaS$_2$

Fig. 18 summarizes the photoluminescence spectra of a slightly Cu-rich CuGaS$_2$ with all the peaks identified. In the course of investigating the CuGaS$_2$ semiconductors in this present work, several well-resolved exciton-related transitions were detected. The bandgap at 10 K is determined as 2.525 eV from the free exciton and its first excited state at 2.496 eV and 2.518 eV, respectively.

In this report, several sub-band edge transitions, were identified as DA transtions interacting with a common shallow donor level at 38±2 meV and shallow acceptors 75±5 meV (A1), 90±3 meV (A2) and 210±5 meV (A3). Metzner et al. have also reported the shallow transitions and defect levels: a shallow donor at 25 meV and two shallow acceptors at 89 meV and 109 meV [30]. In addition, we observe the deeper acceptor A3; the related DA3 transition becomes more intense with higher Cu-content. Botha et al. also reported such defect for slightly Cu-rich CuGaS$_2$: an acceptor 210 meV above the valence band with a donor likely at ~ 53 meV [84]. We have shown that DA1, DA2 and DA3 interact with a common shallow donor ~ 38±3 meV.

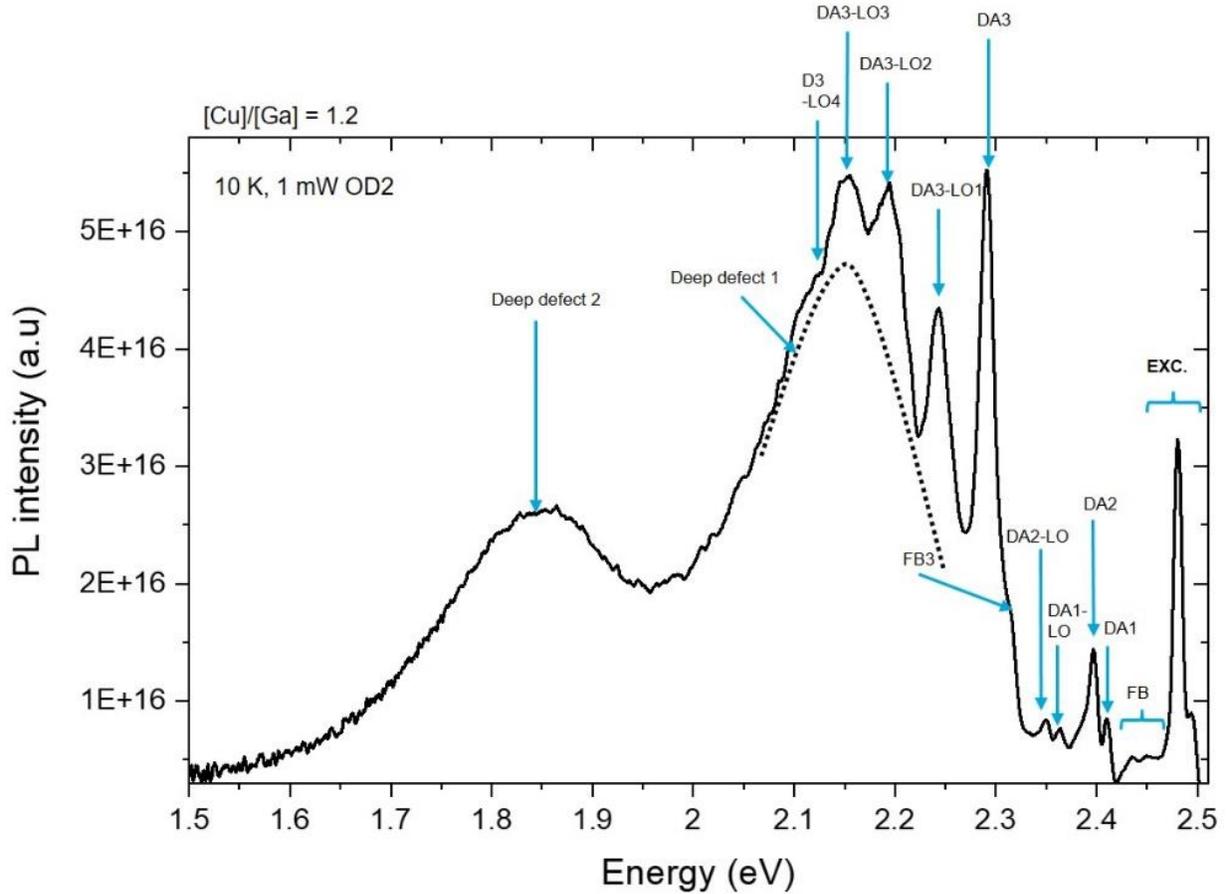

Figure 18: Summary of identified transitions for Cu-rich CuGaS$_2$ at 10 K

Lastly, two broad transitions at 2.15 eV and 1.85 eV were present in all the films. Although both transitions are DA-related, the exact defect levels involved could not be identified, however, both transitions presumably involve broad density of deep states. Transitions involving these defects particularly dominate the PL spectrum of CuGaS$_2$ at room temperature. It is worth mentioning that defect levels identified in this report shares similarity with shallow and deep defects identified in CuGaSe$_2$ [11-14, 19] and CuInS$_2$ [26, 58]. For a detailed comparison see Reference [9]. …

An overview of the transition energies identified from literature and those identified in this work is presented in Fig. 19. It can be observed that different transitions were identified independently by different groups. In this work, within the range of energies investigated, all the different transitions energies separately reported were identified and their composition dependence clarified.

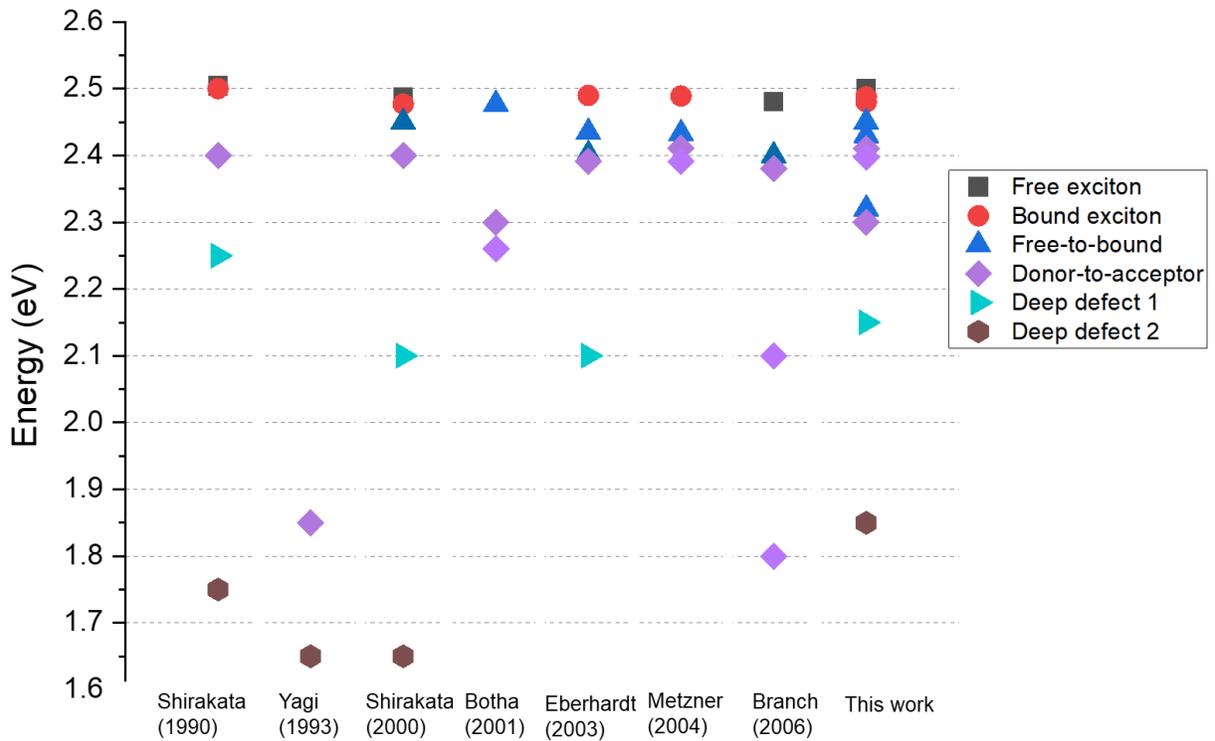

Figure 19: Overview of transition energies of $CuGaS_2$ from literature with transitions identified in this work.

The tentative defect levels on $CuGaS_2$ in Fig. 16 is update and summarized in Fig. 20. The existence of a shallow donor level ~ 35 meV (D1) and shallow acceptors 76 meV (A1), 90 meV (A2) with an additional acceptor deeper level at 210 meV (A3) is observed. Finally, two deep transitions level seem to originate from two broad defect levels deep within the midgap.

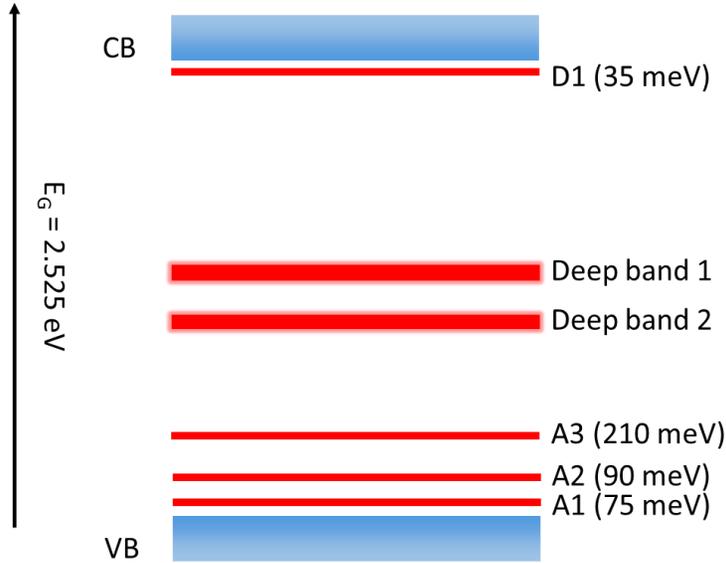

Figure 20: Tentative defect model for CuGaS$_2$ as reported in this work. A shallow donor level (D1) and three shallow acceptor levels (A1, A2 and A3) were identified. Two broad defect levels are also assumed to be involved in transitions in CuGaS$_2$.

## Novel solar cell on CuGaS$_2$

The room temperature bandgap of CuGaS$_2$ is around 2.45 eV. The wide bandgap makes it not interesting for use as a single junction solar cell. Nevertheless, it is also important to understand how the defects of CuGaS$_2$ might influence the electrical properties of a single junction solar cell. The absorber used is a Cu-rich film with a [Cu]/[Ga] ratio of ~1.3. The room temperature PL spectrum of the absorber is shown in Fig. 21. The room temperature spectrum is dominated by the broad transition centered around 1.5 eV as shown in Fig. 21. The device possesses a quasi-Fermi level splitting (QFLS) of 1.68 eV, and consequently a rather large deficit of 0.42 eV compared to the Shockley-Queisser open-circuit voltage ($V_{OC}^{SQ}$) [91] owing to the defects in the material. The QFLS was determined by evaluating the PL quantum efficiency of the absorber, to determine the non-radiative loss or the reducing factor from the ideal $V_{OC}^{SQ}$ [92].

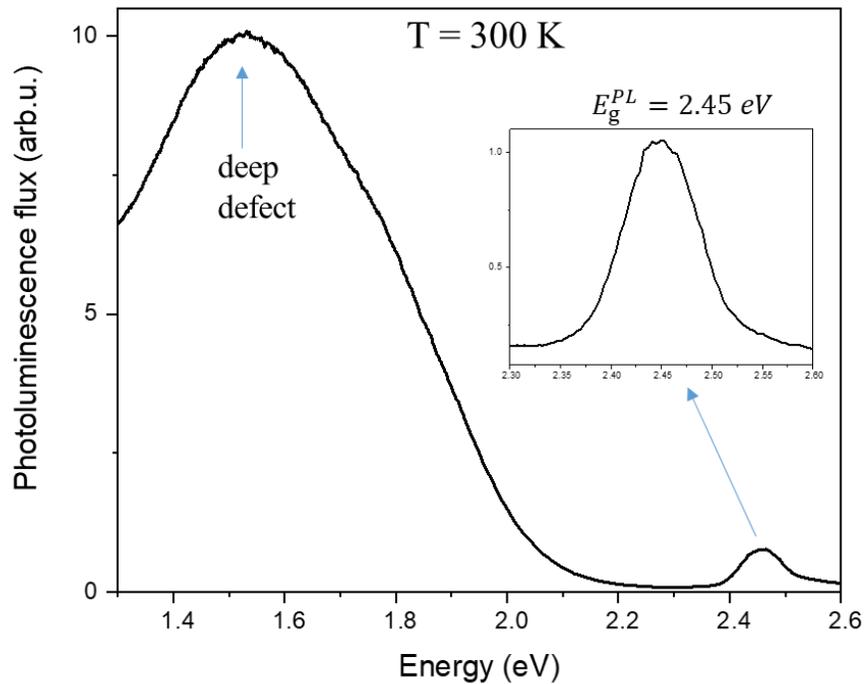

Figure 21: Room temperature photoluminescence spectrum of the CuGaS$_2$ absorber completed into solar cell. In inset is the magnification of the band-to-band transition.

Fig. 22 shows the current density-voltage characteristics of the CuGaS$_2$ device prepared with (Zn,Mg)O buffer layer, Al:ZnMgO sputtered i-layer and an Al:ZnO window layer. The device demonstrated a V$_{OC}$ of 821 mV leading to a very high interface V$_{OC}$ deficit [93] when compared to the quasi-Fermi level splitting. Thus, leading to a power conversion efficiency of mere 1.8%. We speculate that the high interface V$_{OC}$ deficit originates from two factors: (i) from the near interface defects [93] as the device was prepared using the Cu-rich CuGaS$_2$ absorbers, (ii) a negative conduction band offset at the CuGaS$_2$/(Zn,Mg)O interface, due to high conduction band minimum of CuGaS$_2$ and relatively low conduction band minimum of (Zn,Mg)O. While the former limits the V$_{OC}$ by reducing the QFLS near the interface and can be mitigated by doing a chalcogen treatment [94], the later limits the V$_{OC}$ by reducing the QFLS, and requires a buffer that is better matched to the conduction band minimum of the CuGaS$_2$. Nonetheless, This work demonstrates that it is possible to make working solar cells with CuGaS$_2$ though significant efforts are required to achieve useful V$_{OC}$ and power conversion efficiencies.

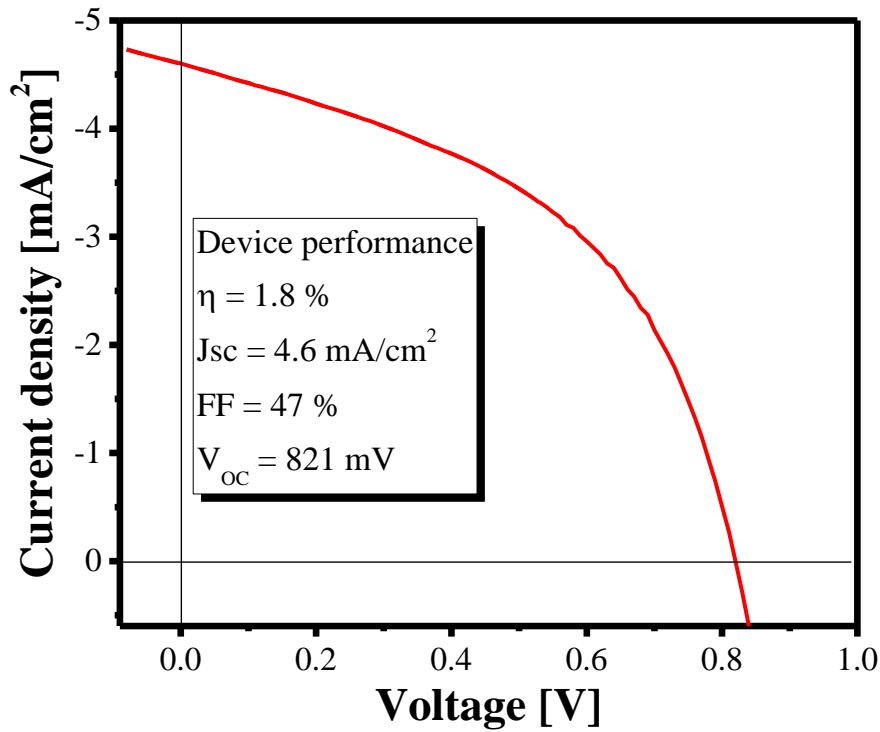

Figure 22: Current density-voltage curve of $CuGaS_2$ device prepared with (Zn,Mg)O buffer layer with Mg/(Mg+Zn) ~ 0.3 atomic percent.

# Acknowledgements

The authors acknowledge this research was funded in whole, or in part, by the Luxembourg National Research Fund (FNR), in the framework of the MASSENA project (grant reference [PRIDE 15/10935404]). For the purpose of open access, the author has applied for a Creative Commons Attributions 4.0 International (CC BY 4.0) license to any Author Accepted Manuscript version arising from this submission. We also acknowledge Dr. Mael Guennou for the Raman spectroscopy on the $CuGaS_2$ film.

# Composition dependence of electronic defects in CuGaS$_2$

Damilola Adeleye*, Mohit Sood, Michele Melchiorre, Alice Debot, Susanne Siebentritt

*Laboratory for Photovoltaics, Department of Physics and Materials Science, University of Luxembourg, Belvaux, L-4422, Luxembourg*

*Corresponding author: damilola.adeleye@uni.lu

# Supplementary information

A simplified single exponential power law described by equation (5.1) is then used as a first step to analyze the dependence of the integrated PL flux on laser excitation intensity [68],

$$I_{PL} \propto \phi^k$$

where $I_{PL}$ is the laser excitation intensity, $\phi$ is the integrated PL flux and $k$ is the power law exponent, taking values $k \leq 1$ for DA and FB, and $1 \leq k \leq 2$ for excitonic transitions.

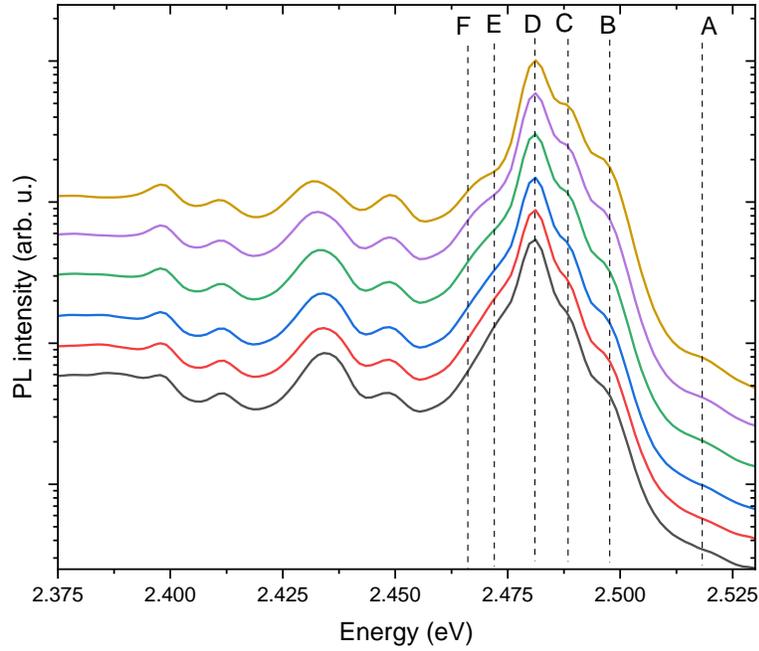

Figure S1: Near band-edge spectra of Cu-rich $CuGaS_2$ measured at 10 K taken at several excitation intensities over three orders of magnitude. The transition peaks are at 2.468 eV (F), ~ 2.474 eV (E), 2.481 eV (D), 2.488 eV (C), 2.496 eV (B) and 2.518 eV (A). The dashed lines highlight the constant energy positions with increasing excitation intensity

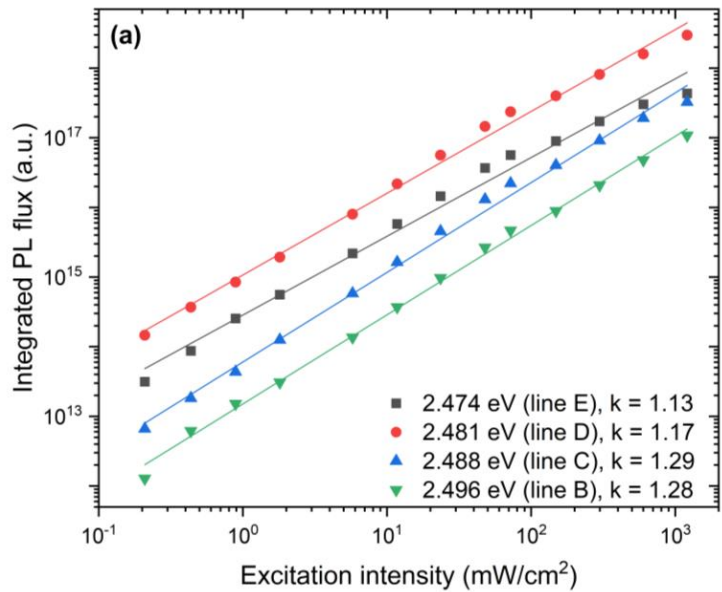

Figure S2: Excitation intensity dependence of integrated PL flux for transitions lines 2.474 eV (E), 2.481 eV (D), 2.488eV (C) and 2.496 eV (B) fitted with a single power law.

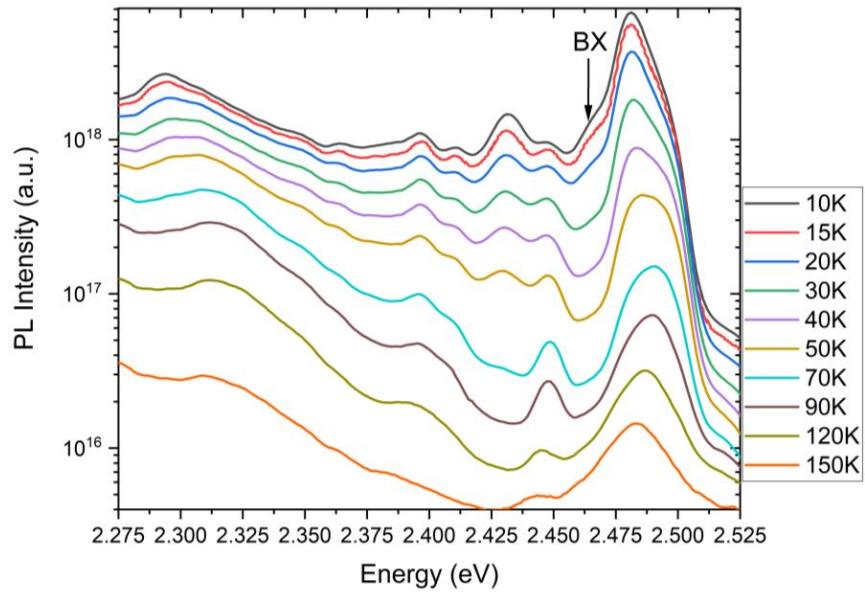

Figure S3: PL spectra of a very Cu-rich film with [Cu]/[Ga] = 2, showing the temperature dependence of band edge and sub-band edge transitions.

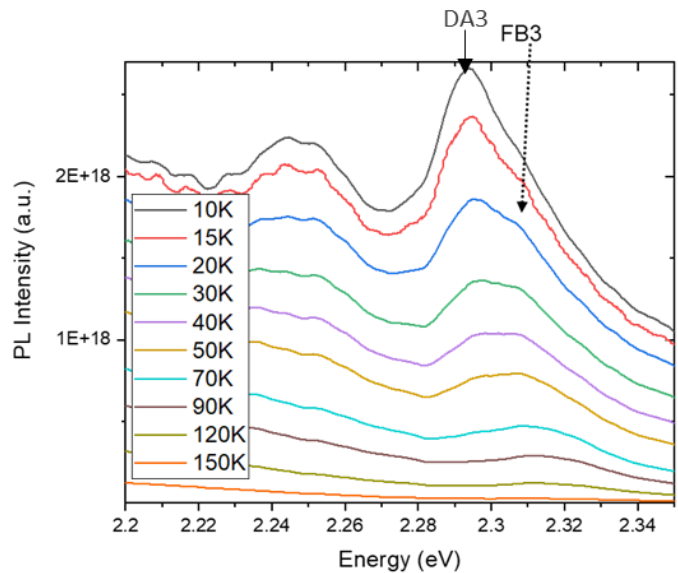

Figure S4: PL spectra of a moderately Cu-rich film with [Cu]/[Ga] ratio of 1.3, showing the temperature dependence DA3 and FB3.

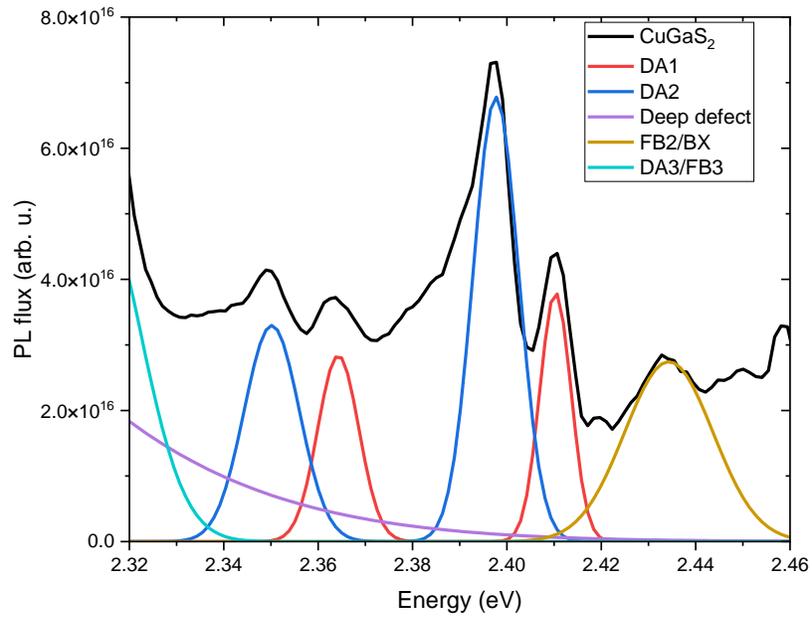

Figure S5: Fitting of DA1 and DA2 and their phonon replicas by a Poisson distribution taking into account deep defects and DA3 and FB3 transitions.